# Rapidly convergent representations for 2D and 3D Green's functions for a linear periodic array of dipole sources

Derek Van Orden and Vitaliy Lomakin

Department of Electrical and Computer Engineering,
University of California, San Diego,
9500 Gilman Dr., La Jolla, CA 92093

## Abstract

Hybrid spectral-spatial representations are introduced to rapidly calculate periodic scalar and dyadic Green's functions of the Helmholtz equation for 2D and 3D configurations with a 1D (linear) periodicity. The presented schemes work seamlessly for any observation location near the array and for any practical array periodicities, including electrically small and large periodicities. The representations are based on the expansion of the periodic Green's functions in terms of the continuous spectral integrals over the transverse (to the array) spectral parameters. To achieve high convergence and numerical efficiency, the introduced integral representations are cast in a hybrid form in terms of (i) a small number of contributions due to sources located around the unit cell of interest, (ii) a small number of symmetric combinations of the Floquet modes, and (iii) an integral evaluated along the steepest descent path (SDP). The SDP integral is regularized by extracting the singular behavior near the saddle point of the integrand and integrating the extracted components in closed form. Efficient quadrature rules are established to evaluate this integral using a small number of quadrature nodes with arbitrary small error for a wide range of structure parameters. Strengths of the introduced approach are demonstrated via extensive numerical examples.

# 1. Introduction

Structures comprising infinite periodic arrays of elements find many applications in physics and engineering including microwave and optical filters, printed and leaky-wave antennas, lasers, and waveguides. An efficient method to compute electromagnetic fields in such structures is to use the integral approach [1], in which fields are found in terms of a superposition integral (spatial convolution) of a source distribution, given within a unit cell, and a periodic Green's function (PGF), which incorporates the periodic boundary conditions. Efficient methods for computing PGFs are important to enable the use of integral-equation methods, to compute fields radiated and scattered from periodic configurations, and to study their dispersion properties.

It is well known that representations of PGFs via direct spatial summations are extremely slowly convergent, and diverge when the periodicity is defined with a complex phase shift between elements [1]. The convergence can be improved by using spectral summations in terms of Floquet series. However Floquet series representations suffer from slow convergence in the important case where the observation point resides near a periodic structure. For 3D problems with 1D (linear) periodicity the Floquet series are divergent along the periodicity axis.

Several approaches have been proposed to accelerate the computation of PGFs. Some are based on spectral and spatial formulations with Poisson's, Kummer's, and Shank's transformations [1-5], which accelerate the original spatial and spectral series representations but do not lead to exponentially rapid convergence and may be slow in various situations [1, 2, 6]. The Ewald approach, in contrast, using Ewald and Poisson's transformations, leads to exponentially convergent PGF series representations. However, it requires double summations, can suffer high-frequency breakdown, and involves choosing a proper splitting parameter, which may be not straightforward to implement [1, 2, 6-16]. Moreover, the Ewald approach has never been extended to 3D dyadic Green's functions. Veysoglu's transformation [1, 2, 6, 17] and perfectly matched layer (PML) approaches [18, 19] also can be used to lead to exponential convergence. However, as explained below, these approaches are not applicable to complex phase shifted conditions and may suffer from low-frequency breakdown.

In this paper we introduce a novel method for computing scalar and dyadic PGFs for 2D and 3D problems comprising a 1D periodicity with phase shifted conditions. The method is based on the alternative Green's function spectral representation formalism [20] and involves the transverse (to the periodicity axis) spectral expansions with the longitudinal (along the axis) spectral 1D PGF available in closed form. These representations lead to exponential computational convergence for any location near the array axis and for any practical periodicities, including electrically small and large periodicities. Moreover, the introduced series have a clear physical interpretation, being expressed in terms of a small number of direct field contributions, a small number of Floquet modes, and a rapidly convergent integral representing the combined contribution of the remaining sources.

The paper is organized as follows. Section 2 presents the problem formulation. Section 3 is a mathematical presentation of the resolvent formalism, which is then used in section 4 to derive alternative spectral representations for PGFs in terms of the longitudinal representation (conventional Floquet mode expansion) and transverse representation (novel spectral approach). The latter representation is used in Sec. 5 with related Appendices A and B to develop highly efficient (exponentially accurate and rapid) schemes for computing scalar 2D and 3D PGFs. Section 6 outlines how the procedure in Sec. 5 can be extended to rapidly compute the dyadic 2D and 3D PGFs. Section 7 presents an extensive numerical study supporting the theoretical results in Secs. 3-6. Section 8 draws conclusions and summarizes findings of the paper.

## 2. Problem formulation

Consider a 1D periodic array of identical electric dipole sources in free space. The dipoles are arranged along the $x$ axis with spacing $L$ and directed along a general unit vector $\hat{\mathbf{p}}$ (Fig. 1). These are linear sources for 2D configurations and point sources for 3D problems. The sources have a linear phase shift $e^{-jk_{x0}nL}$ determined by the (generally complex) parameter $k_{x0}$, where $n$ is an integer counting the sources such that the source with $n = 0$ lies at the origin of the coordinate system (Fig. 1). The time dependence $e^{j\omega t}$, with $\omega$ being the angular frequency, is assumed and suppressed everywhere in the paper.

Electromagnetic fields radiated by this array can be found using dyadic PGFs. Specifically, the electric field radiated by the array in Fig. 1 can be found from

$$\mathbf{E}^{2D,3D} = \underline{\underline{\mathbf{G}}}^{2D,3D} \cdot Il\hat{\mathbf{p}}. \tag{1}$$

where $Il$ is the current moment of each dipole. (Here and in the rest of the paper the superscript $2D, 3D$ denotes a function for 2D or 3D cases.) The 2D and 3D dyadic PGFs $\underline{\underline{\mathbf{G}}}^{2D,3D}$ are given by

$$\underline{\underline{\mathbf{G}}}^{2D,3D}(\mathbf{r}) = -\frac{j\eta}{k}(k^2\underline{\underline{\mathbf{I}}} + \nabla\nabla)G^{2D,3D}. \tag{2}$$

Here, $\eta$ is the free-space characteristic impedance, $k$ is the free-space wavenumber, and $G^{2D}$ and $G^{3D}$ are respectively 2D and 3D scalar PGFs that are defined via the superposition principle

$$G^{2D,3D}(\mathbf{r}) = \sum_{n=-\infty}^{\infty} e^{-jk_{x0}nL} G_0^{2D,3D}(k, R_n),$$

$$G_0^{2D}(k, R_n) = \frac{1}{4j} H_0^{(2)}(kR_n), \qquad G_0^{3D}(k, R_n) = \frac{e^{-jkR_n}}{4\pi R_n}. \tag{3}$$

where $G_0^{2D}$ and $G_0^{3D}$, the scalar 2D and 3D free-space Green's functions due to a single isolated source, are summed over all the sources in the array. Here, $R_n = |\mathbf{r} - \mathbf{r}_n|$ is the distance from the observer located at $\mathbf{r}$ to the $n$-th source at $\mathbf{r}_n$. The scalar PGF $G^{2D}$ is a function of the two spatial variables $x$ and $z$, while $G^{3D}$ depends only on $x$ and $\rho = (y^2 + z^2)^{1/2}$. Clearly, the scalar (and thus dyadic) PGFs exhibit the phase shifted periodicity property with respect to the periodicity $L$

$$G^{2D,3D}(\mathbf{r}+nL\hat{\mathbf{x}}) = e^{-jk_{x0}nL}G^{2D,3D}(\mathbf{r}),$$
$$\underline{\underline{\mathbf{G}}}^{2D,3D}(\mathbf{r}+nL\hat{\mathbf{x}}) = e^{-jk_{x0}nL}\underline{\underline{\mathbf{G}}}^{2D,3D}(\mathbf{r}). \quad (4)$$

The conditions in Eq. (4) can be regarded not only as a property of the PGFs but also as boundary conditions, which relate the fields on the two boundaries of each unit cell.

For lossless media the series in Eq. (3) converges extremely slowly. Therefore, more efficient representations for constructing the PGFs are to be developed.

## 3. Spectral resolvent formalism

The scalar PGFs in Eq. (3) can be found by separation of variables into the transverse ($z$ or $\rho$) and longitudinal ($x$) variables. To this end, the PGFs are expanded over a series of basis functions constituting an orthogonal basis for the problem either in the longitudinal or transverse direction. The expansion over the longitudinal spectrum leads to the well known Floquet mode representation. The expansion over the transverse spectrum has never been exploited for PGFs and is studied below.

More generally, the PGFs can be expressed as a contour integration of a product of spectral 1D or 2D Green's functions as defined in the framework of the resolvent approach by Felsen and Marcuvitz [20]. The PGFs are represented as

$$G^{2D}(\mathbf{r}) = \frac{1}{2\pi j}\oint d\lambda_x g_x(x,\lambda_x)g_z(z,\lambda_z(\lambda_x)) = \frac{1}{2\pi j}\oint d\lambda_z g_z(z,\lambda_z)g_x(x,\lambda_x(\lambda_z)),$$
$$G^{3D}(\mathbf{r}) = \frac{1}{2\pi j}\oint d\lambda_x g_x(x,\lambda_x)g_\rho(\rho,\lambda_\rho(\lambda_x)) = \frac{1}{2\pi j}\oint d\lambda_\rho g_\rho(\rho,\lambda_\rho)g_x(x,\lambda_x(\lambda_\rho)). \quad (5)$$

Here, $\lambda_x$ and $\lambda_{z,\rho}$ are spectral variables associated with the longitudinal and transverse separable variables $x$ and $z$ or $\rho$, respectively. These spectral parameters are related via $-\lambda_x - \lambda_{z,\rho} = k^2$.

In Eqs. (5), the functions $g_z(z,\lambda_z)$ and $g_\rho(\rho,\lambda_\rho)$ are 1D and 2D PGFs for the spectral problem in the transverse direction. They are given by

$$g_z(z, \lambda_z) = \frac{e^{-j\sqrt{-\lambda_z}|z|}}{2j\sqrt{-\lambda_z}},$$

$$g_\rho(\rho, \lambda_\rho) = \frac{1}{4j} H_0^{(2)}\left(\sqrt{-\lambda_\rho}\rho\right). \tag{6}$$

These functions have branch points at $\lambda_{z,\rho} = 0$, and the associated branch cuts are chosen along the positive real axis (see Fig. 2(a)). The top Riemann sheet, in which integrations are carried out, is chosen such that $\text{Im}\{(-\lambda_{z,\rho})^{1/2}\} < 0$.

In both equations of (5), the function $g_x(x, \lambda_x)$ is the 1D PGF for the spectral problem in the $x$ direction. It is given by

$$g_x(x, \lambda_x) = \frac{1}{2j\sqrt{-\lambda_x}} \left( \frac{e^{-j\sqrt{-\lambda_x}|x|}}{1 - e^{-j(\sqrt{-\lambda_x} - k_{x0})L}} - \frac{e^{j\sqrt{-\lambda_x}|x|}}{1 - e^{j(\sqrt{-\lambda_x} + k_{x0})L}} \right). \tag{7}$$

for $|x| < L$. It can be found by summing up contributions from all point sources for the 1D spectral problem in the $x$ direction

$$g_x(x, \lambda_x) = \left(2j(-\lambda_x)^{1/2}\right)^{-1} \sum_{n=-\infty}^{\infty} e^{-jk_{x0}nL} e^{-j\sqrt{-\lambda_x}|x-nL|}. \tag{8}$$

The spectral PGF $g_x(x, \lambda_x)$ satisfies the 1D spectral Helmholtz equation $(d^2/dx^2 - \lambda_x)g_x(x, \lambda_x) = -\delta(x)$ in the $n = 0$ unit cell with the boundary conditions

$$g_x(x + L, \lambda_x) = e^{-jk_{x0}L} g_x(x, \lambda_x). \tag{9}$$

The PGF $g_x(x, \lambda_x)$ has an infinite number of poles in the complex $\lambda_x$ plane but it does not have any branch points and branch cuts.

In Eqs. (5), the integration contours in the complex $\lambda_x$ plane enclose the singularities of $g_x(x, \lambda_x)$ but not those of $g_z(z, \lambda_z(\lambda_x))$ and $g_\rho(\rho, \lambda_\rho(\lambda_x))$. Similarly, the integrations contours over $\lambda_z$ and $\lambda_\rho$ enclose the singularities of $g_z(z, \lambda_z)$ and $g_\rho(\rho, \lambda_\rho)$ but not those of $g_x(x, \lambda_x(\lambda_{z,\rho}))$ [20].

The representations of Eqs. (5) in terms of the longitudinal and transverse spectral parameters yield identical results. The choice between them is mostly one of computational convenience and convenience of representation. The two representation choices are analyzed next.

4. Alternative spectral representations

The formalism in Sec. 3 can be used to develop two alternative representations for the 2D and 3D PGFs by using the integrations over the longitudinal and transverse spectral variables in Eq. (5). Section 4.1 presents the longitudinal (Floquet mode) spectrum representations. Section 4.2 introduces the novel transverse spectrum representation.

4.1 Longitudinal (Floquet mode) representation

In this representation, the integration over the longitudinal spectral parameter $\lambda_x$ in Eq. (5) is carried out. Because $g_x(x, \lambda_x)$ has no branch cuts in this complex plane, the only contribution to the integral comes from the residues of its infinite number of poles at $\lambda_{xm} = -\left(k_{x0} + 2\pi m/L\right)^2$ (Fig. 2(b)). Each residue contribution is identified as the contribution of the corresponding Floquet (diffraction) mode. (Here and in the rest of the paper the integer $m$ will be used to count the poles of $g_x$ and corresponding Floquet modes of the array.) The resulting sum of these residues leads to the conventional Floquet mode expansions

$$G^{2D}(\mathbf{r}) = \frac{1}{L} \sum_{m=-\infty}^{\infty} \frac{e^{-jk_{xm}x} e^{-jk_{zm}|z|}}{2jk_{zm}},$$

$$G^{3D}(\mathbf{r}) = \frac{1}{L} \sum_{m=-\infty}^{\infty} e^{-jk_{xm}x} \frac{1}{4j} H_0^{(2)}(k_{\rho m}\rho), \tag{10}$$

where $k_{xm} = k_{x0} + 2\pi m/L$ and $k_{\{z,\rho\}m} = (k^2 - k_{xm}^2)^{1/2}$ are referred to as the longitudinal and transverse Floquet mode wavenumbers, respectively. The choice of the square root for $k_{\{z,\rho\}m}$ depends on the application. In many cases, e.g. when the interest is to obtain the scattering coefficients of a periodic structure, the square root is chosen so that $k_{\{z,\rho\}m}$ is on the upper Riemann sheet with $\text{Im}\{k_{\{z,\rho\}m}\} \leq 0$ for all $m$. In other cases, such as the analysis of complex dispersion relations of leaky-wave antennas, one may choose $\text{Im}\{k_{\{z,\rho\}m}\} > 0$ for a finite number of Floquet modes $m$.

The sum (10) converges exponentially fast when $|z|, \rho$ are large with the number of terms required to achieve a high accuracy being greater than $L/(2\pi |\{z,\rho\}|)$. On the other hand, as $|\rho|, |z| \to 0$, the summations in (10) converges extremely slowly. However, it is this latter case, where the fields are observed along the axis of the sources, that is often of interest. Therefore, alternative rapidly convergent representations for $G^{2D}$ and $G^{3D}$ are to be developed.

It is noted that the Floquet mode representation (10) can be obtained in other ways. In this paper, the derivation of (10) exemplifies the use of the resolvent formalism and emphasizes an intimate relation between the Floquet mode spectrum and the 1D spectral PGF $g_x(x, \lambda_x)$. This approach also demonstrates the relations between the longitudinal and transverse representations.

4.2 Transverse representation

Integration over the transverse spectral parameters $\lambda_z$ and $\lambda_\rho$ in Eq. (5) yields the transverse representation. The only contribution to the integral comes from integrating around the branch cut along the positive part of the real axis (Fig. 2(a)). Changing variables from the spectral variable $\lambda_{z,\rho}$ to $k_{z,\rho} = (-\lambda_{z,\rho})^{1/2}$ and exploiting the fact that $g_x$ is a symmetric function of $k_{z,\rho}$ leads to an alternative representation for the 2D and 3D PGFs in terms of the transverse spectrum expansion

$$G^{2D}(\mathbf{r}) = \frac{1}{\pi} \int_0^\infty dk_z \cos(k_z z) g_x(x, k_x(k_z)),$$

$$G^{3D}(\mathbf{r}) = \frac{1}{2\pi} \int_0^\infty dk_\rho k_\rho J_0(k_\rho \rho) g_x(x, k_x(k_\rho)).$$

(11)

Here, $J_0(k_\rho\rho)$ is the zeroth-order Bessel function, $k_{z,\rho}$ is identified as the wavenumber in the transverse direction, $k_x = (-\lambda_{z,\rho})^{1/2} = (k^2 - k_{z,\rho}^2)^{1/2}$ is identified as the wavenumber in the $x$ direction, and $g_x(x,k_x)$ is the longitudinal 1D PGF that is redefined in terms of the longitudinal wavenumber $k_x$

$$g_x(x,k_x) = \frac{1}{2jk_x}\left(\frac{e^{-jk_x|x|}}{1-e^{-j(k_x-k_{x0})L}} - \frac{e^{jk_x|x|}}{1-e^{j(k_x+k_{x0})L}}\right). \quad (12)$$

As clear from the discussion after Eq. (10), the spectral PGF $g_x(x,k_x)$ has poles in the complex $k_{z,\rho}$ plane at $k_{\{z,\rho\}m} = \left(k^2 - (k_{x0} + 2\pi m/L)^2\right)^{1/2}$, which correspond to the Floquet mode wavenumbers $k_{xm} = k_{x0} + 2\pi m/L$. For lossless configurations and purely real $k_{x0}$, these poles form pairs with opposite sign on either the imaginary or real axis of $k_z$. For observation points within the $n=0$ unit cell, i.e. for $|x|<L$, $g_x(x;k_x)$ decays exponentially fast for large $|k_{z,\rho}|$. This decay leads to an exponentially fast convergence of the representation (11) even for the case $|z|=\rho=0$, provided that $x\neq 0$. For this reason the longitudinal representation is the better choice for finding the fields at or near the axis of the sources.

The result of the transverse representation in (11) is identical to that in (10) if all square roots in (10) are defined on the upper Riemann sheet, i.e. if $\text{Im}\{k_{\{z,\rho\}m}\} \leq 0$. To define the PGFs on the lower Riemann sheet associated with a wavenumber $k_{\{z,\rho\}M}$ defined in (10) via $\text{Im}\{k_{\{z,\rho\}M}\} > 0$, the expression in (11) is modified as

$$G^{2D}(\mathbf{r}) \to G^{2D}(\mathbf{r}) - \frac{2}{L}\frac{\cos(k_{zM}z)e^{-jk_{xM}x}}{2jk_{zM}},$$
$$G^{3D}(\mathbf{r}) \to G^{3D}(\mathbf{r}) - \frac{2}{L}\frac{j}{4}J_0(k_{\rho M}\rho)e^{-jk_{xM}x}. \quad (13)$$

Here, $G^{2D,3D}(\mathbf{r})$ in the left and right-hand sides are PGFs defined on the lower and upper Riemann sheet of $k_{\{z,\rho\}M}$, respectively. This expression can be obtained by taking the difference between $G^{2D,3D}(\mathbf{r})$ on the lower and upper Riemann sheets. The rest of the analysis in the paper assumes that the PGFs are defined on the upper Riemann sheets of $k_{\{z,\rho\}m}$ for all $m$. The PGFs on the lower Riemann sheets can be easily obtained via (13). It is also noted that in all practical situations only a small number of lower Riemann sheets (typically 1 or 2) is required to be considered.

Compared to the Floquet mode expansions (10), the representations in (11) are better suited for computing the PGFs near the array. However, they still may suffer from problems in their numerical implementations. For the case $\mathbf{r} \to 0$ (i.e. for $x, z, \rho \to 0$) the representations in (11) are slowly convergent with convergence rate similar to that of (10). In addition, the integrands in (11) may have poles on the real (integration) axis. Furthermore, for large $kL$, i.e. in the high-frequency regime, the integrands are highly oscillatory, which can complicate their direct evaluation. These potential problems are resolved in Sec. 5.

## 5. Fast computation of the scalar periodic Green's functions

This section presents a procedure that uses the transverse spectral expansions in Sec. 4.2 to derive rapidly convergent representations for the 2D and 3D PGFs. The representations are valid for any source-observer location near the array and can be easily implemented in a computer code. The procedure involves manipulating the integrands in (11) to make them slowly varying and thus easily integrable. Section 5.1 presents such a rapidly convergent representation. Section 5.2 presents an efficient numerical implementation of the representation in Sec. 5.1.

5.1 Regularized transverse representation

The convergence of the integrals (11) can be improved by explicitly extracting the contribution of a certain number of sources in and around the unit cell of interest and evaluating them directly using the spatial PGFs in Eq. (3). The remaining infinite sources may be evaluated using the transverse representation (11). To formulate this, the longitudinal spectral PGF $g_x(x, k_x)$ in (8) can be rewritten as

$$g_x(x,k_x) = \left( \sum_{n=-N_d}^{N_d} \frac{e^{-jk_{x0}nL} e^{-jk_x|x-nL|}}{2jk_x} \right) + \frac{e^{-jk_xL(N_d+1)}}{2jk_x} \left( \frac{e^{jk_{x0}L(N_d+1)} e^{-jk_xx}}{1-e^{-j(k_x-k_{x0})L}} + \frac{e^{-jk_{x0}L(N_d+1)} e^{jk_xx}}{1-e^{-j(k_x+k_{x0})L}} \right). \quad (14)$$

Here, the first term in the right-hand side is a summation of free-space longitudinal spectral 1D PGF, taken over $2N_d+1$ sources residing at locations $x=nL$ around the origin. The second term in the right-hand of (14), which comprises the contribution from the remaining infinite number of sources, decays exponentially fast for large $k_{z,\rho}$ regardless of the observation location $x$ within the $n=0$ (zeroth) unit cell (and even for $x=0$).

To avoid the possible oscillatory behavior of the integrands in (11) for large $L$, the integration path can be deformed from the original path along the real axis to the to the steepest descent path (SDP) [20, 21] (see Fig. 3). In general, the SDP path should pass through a saddle point whose position in the complex $k_{z,\rho}$ plane depends on $N_dL$, $x$, and $z$. Here, to obtain a robust scheme for any $\mathbf{r}$ withunit cell, it is assumed that $N_dL > |\mathbf{r}|$ and the saddle point is given by $k_{z,\rho}=0$. Residues at poles that are crossed while deforming the original integration path to the SDP must then be taken into account. To this end, the scalar PGFs can be written as

$$G^{2D}(\mathbf{r}) = \sum_{n=-N_d}^{N_d} e^{-jk_{x0}nL} \frac{1}{4j} H_0^{(2)}(kR_n) + \frac{2}{L} \sum_{m=1}^{N_p} \frac{\cos(k_{zm}z) e^{-jk_{xm}x}}{2jk_{zm}} + G_{SDP}^{2D}(\mathbf{r}),$$

$$G^{3D}(\mathbf{r}) = \sum_{n=-N_d}^{N_d} e^{-jk_{x0}nL} \frac{e^{-jkR_n}}{4\pi R_n} + \frac{2}{L} \sum_{m=1}^{N_p} \frac{j}{4} J_0(k_{\rho m}\rho) e^{-jk_{xm}x} + G_{SDP}^{3D}(\mathbf{r}). \quad (15)$$

Here, the first summations are obtained by using the first terms in the right-hand side of (14) in (11). These summations represent contributions due to $N_d+1$ sources located in and around the origin (Fig. 1), and can also be found by taking the sum in (3) over these sources. The second summations in (15) represent contributions of $N_p$ poles crossed while the deforming the original integration path to the SDP. Comparing with the Floquet mode representation (10), these pole contributions are identified as combinations of the Floquet modes with symmetric behavior with respect to the direction transverse to the array axis ($z$ and $\rho$ directions). The functions $G_{SDP}^{2D}(\mathbf{r})$ and $G_{SDP}^{3D}(\mathbf{r})$ in (15) are SDP integrals given by

$$G_{SDP}^{2D}(\mathbf{r}) = \frac{1}{\pi} \int_{SDP} \left( \frac{e^{jk_{x0}L(N_d+1)} e^{-jk_x x}}{1 - e^{-j(k_x - k_{x0})L}} + \frac{e^{-jk_{x0}L(N_d+1)} e^{jk_x x}}{1 - e^{-j(k_x + k_{x0})L}} \right) \frac{e^{-jk_x L(N_d+1)} \cos(k_z z)}{2 jk_x} dk_z,$$

$$G_{SDP}^{3D}(\mathbf{r}) = \frac{1}{2\pi} \int_{SDP} \left( \frac{e^{jk_{x0}L(N_d+1)} e^{-jk_x x}}{1 - e^{-j(k_x - k_{x0})L}} + \frac{e^{-jk_{x0}L(N_d+1)} e^{jk_x x}}{1 - e^{-j(k_x + k_{x0})L}} \right) \frac{e^{-jk_x L(N_d+1)} J_0(k_\rho \rho)}{2 jk_x} k_\rho dk_\rho.$$

(16)

These functions represent the contribution of the remaining sources. Unlike the integrands in Eq. (11), the integrands in Eq. (16) have branch points associated with the square root $k_x = (k^2 - k_{z,\rho}^2)^{1/2}$, chosen such that $\text{Im}\{k_x\} < 0$. These branch points, however, can be eliminated by a change of variables.

The integrands in (16) decay very rapidly with an increase of $k_{z,\rho}$. To show this explicitly the SDP integral can be formulated by making a change in variable from $k_{z,\rho}$ to the dimensionless variable $s$, defined via $k_x = k(1 - js^2)$. The SDP integrals on the right-hand side of Eq. (16) can be rewritten as

$$G_{SDP}^{2D}(\mathbf{r}) = \int_0^\infty f^{2D}(s) e^{-\alpha s^2} \frac{ds}{\sqrt{s^2 + 2j}},$$

$$G_{SDP}^{3D}(\mathbf{r}) = \int_0^\infty f^{3D}(s) e^{-\alpha s^2} s \, ds,$$

(17)

where $\alpha = kL(N_d + 1)$ and

$$f^{2D}(s) = \frac{1}{\pi} \left( \frac{e^{jk_{x0}L(N_d+1)} e^{-kx(j+s^2)}}{1 - e^{jk_{x0}L} e^{-kL(j+s^2)}} + \frac{e^{-jk_{x0}L(N_d+1)} e^{kx(j+s^2)}}{1 - e^{-jk_{x0}L} e^{-kL(j+s^2)}} \right) \cos(kz\, s\sqrt{s^2 + 2j}),$$

$$f^{3D}(s) = \frac{1}{2\pi} \left( \frac{e^{jk_{x0}L(N_d+1)} e^{-kx(j+s^2)}}{1 - e^{jk_{x0}L} e^{-kL(j+s^2)}} + \frac{e^{-jk_{x0}L(N_d+1)} e^{kx(j+s^2)}}{1 - e^{-jk_{x0}L} e^{-kL(j+s^2)}} \right) J_0\left(k\rho\, s\sqrt{s^2 + 2j}\right).$$

(18)

The factor of $e^{-\alpha s^2}$ in each integral ensures a rapid decay of the integrand and this decay is more significant for large $N_d$. Typically, a small $N_d$ is adequate to render sufficient convergence, and for most practical problems only a small number of poles $N_p$ need to be taken into account. An important difference between the integrands in (17) is that the integrand of $G_{SDP}^{2D}$ has a branch cut $(s^2 + 2j)^{1/2}$, which appears due to the variable transformation from $k_{z,\rho}$ to $s$, whereas the integrand of $G_{SDP}^{3D}$ does not. As explained in Sec. 5.2, the presence of the branch cut may lead to a low-frequency breakdown for the 2D case and requires modifications in handing the low-frequency regime ($L \ll \lambda$). The absence of the branch cut in the integrand of $G_{SDP}^{3D}$ in (16) is due to presence of the $k_\rho$ factor in the integrand for this case. Robust numerical implementations for the integrations in (17) are given in Sec. 5.2.

5.2 Robust numerical implementation of the transverse representation

When the poles in the integrands in Eq. (18) reside near the saddle point $s = 0$, the integrands vary rapidly along the SDP near these poles. Such situations are encountered when the poles are within the contributing zone of the integrand, i.e. when $|s_m| < \Omega_p / \alpha^{1/2}$ with $\Omega_p$ being a parameter of order $O(1)$. This occurs in the low-frequency regime ($L \ll \lambda$) where $\alpha \ll 1$ and/or when the wavenumber of one of the Floquet modes is close to the free-space wavenumber, i.e. when $k_{x0} + 2\pi m/L \approx \pm k$. In the latter case the poles $s_m$ can be very close to $s = 0$ and $|s_m| < \Omega_p / \alpha^{1/2}$ even for a large $\alpha$. In such cases, the rate of the integrand variations due to the presence of the poles may be much greater than that due to other terms. To smooth these variations the residues of the integrands at these poles can be explicitly extracted and integrated in a closed form. It is noted that the symmetry of $g_x(x, k_x)$ with respect to $k_{z,\rho}$ and $s$ leads to the existence of pairs of poles that are symmetric about the origin in the complex $s$ plane. Denote as $\pm s_m$ the $m$ th pair of poles that reside near the saddle point $s = 0$ and assume that there are a total (typically small) number of $\tilde{N}_p$ pole pairs to account for.

One can then rewrite the SDP integrals in (17) as

$$G_{SDP}^{2D}(\mathbf{r}) = \int_0^\infty \tilde{f}^{2D}(s) e^{-\alpha s^2} \frac{ds}{\sqrt{s^2 + 2j}} + \sum_{n=1}^{\tilde{N}_p} 2 s_m \text{Res}\{f^{2D}(s_m)\} I_m^{2D},$$

$$G_{SDP}^{3D}(\mathbf{r}) = \int_0^\infty \tilde{f}^{3D}(s) e^{-\alpha s^2} s\, ds + \sum_{n=1}^{\tilde{N}_p} 2 s_m \text{Res}\{f^{3D}(s_m)\} I_m^{3D}.$$

(19)

Here,

$$\tilde{f}^{2D,3D}(s) = f^{2D,3D}(s) - \sum_{m=1}^{\tilde{N}_p} \frac{2 s_m \text{Res}\{f^{2D}(s_m)\}}{s^2 - s_m^2} \tag{20}$$

are non-resonant functions that have no singularities near the integration path. The integrals $I_m^{2D}$ and $I_m^{3D}$ in (19) are given by

$$I_m^{2D} = \int_0^\infty \frac{e^{-\alpha s^2}}{s^2 - s_m^2} \frac{ds}{\sqrt{s^2 + 2j}} = \frac{\pi e^{-\alpha s_m^2} \text{erfc}(j s_m \alpha\, \text{sign}(j s_m))}{2\sqrt{s_m^2 + 2j}\,\text{sign}(j s_m)} + \int_0^\infty \frac{e^{-\alpha s^2}}{s^2 - s_m^2}\left(\frac{1}{\sqrt{s^2 + 2j}} - \frac{1}{\sqrt{s_m^2 + 2j}}\right) ds,$$

$$I_m^{3D} = \int_0^\infty \frac{e^{-\alpha s^2}}{s^2 - s_m^2} s\, ds = \frac{1}{2} e^{-\alpha s_m^2} \text{Ei}(-\alpha s_m^2),$$

(21)

where erfc is the complementary error function, and Ei is the exponential integral function [22]. The integrals for $I_m^{2D}$ can be rapidly evaluated numerically using various adaptive integration methods. The integrand of the first integral for $I_m^{2D}$ has a pole and a branch-cut singularity, whereas the integrand of the second integral has only the branch-cut singularity, which allows for a simpler numerical evaluation. It is important to mention that the integrals and special functions in (21) do not depend on the observation location and need to be evaluated only once for a given problem. Therefore, they do not contribute significantly to the overall computational cost.

The first integration terms in the right-hand side of (19) can be evaluated using various quadrature rules, e.g. those developed in Appendix A and B. The resulting SDP integrals are given by

$$G_{SDP}^{2D,3D}(\mathbf{r}) = \sum_{q=1}^{N_q} \tilde{f}^{2D,3D}(s_q^{2D,3D})e^{-\alpha(s_q^{2D,3D})^2} w_q^{2D,3D} + \sum_{n=1}^{\tilde{N}_p} 2s_m \text{Res}\{f^{2D,3D}(s_m)\}I_m^{2D,3D} \, , \quad (22)$$

where $s_q^{2D,3D}$ are quadrature nodes for 2D and 3D problems given in Appendix A and B, $w_q^{2D,3D}$ are corresponding quadrature weights, and $N_q$ is the number of quadrature nodes (and weights). For the 2D case, the quadrature rules are defined differently for high-frequency ($\alpha \gg 1$, i.e. $L \sim \lambda$ or $L \gg \lambda$) and low-frequency ($\alpha \ll 1$, i.e. $\lambda \ll L$) regimes. For the high frequency regime ($\alpha \gg 1$), a simple equally-spaced node quadrature rule is very efficient (Appendix A.1). For the low-frequency regime, the square root $(s^2+2j)^{-1/2}$ varies much faster than the exponential term $\exp(-\alpha s^2)$, and requires a special treatment. As shown in Appendix A.2, the square-root term can be expanded over a series of cosine functions and this series can be truncated to result in a quadrature rule with a small number of quadrature nodes for any $\alpha$. For the 3D case, an efficient quadrature rule is obtained in Appendix B based on the integrand expansion over Bessel function series. This rule applies to any periodicity (electrically small and large) without modification because the 3D integrand in (19) does not have a branch-cut singularity.

Representations in Eq. (15) with the SDP contributions in (22) are highly efficient for computing the 2D and 3D PGF for any observation location near the structure (whether for $z = \rho = 0$, or for $\mathbf{r} = 0$, or for $z, \rho$ of several $\lambda$) and for any practical periodicities $L$ (whether electrically small with $L \ll \lambda$, intermediate with $L \sim \lambda$, and electrically large $L \gg \lambda$). The numbers of quadrature nodes $N_q$, pole contributions $N_p$, extracted pole terms $\tilde{N}_p$, and direct components $N_d$ combined can be small to achieve any accuracy up to double precision.

## 6. Fast computation of the dyadic periodic Green's functions

Having established a procedure for calculating the scalar PGFs $G^{2D,3D}$, one may apply the dyadic operator in Eq. (2) to find the full dyadic PGFs $\underline{\underline{\mathbf{G}}}^{2D,3D}$. The dyadic operator, involving derivatives with respect to spatial coordinates, is brought inside the integrals in Eq. (11)

$$\underline{\underline{\mathbf{G}}}^{2D}(\mathbf{r}) = -\frac{j\eta}{\pi k}\int_0^\infty dk_z (k^2\underline{\underline{\mathbf{I}}} + \nabla\nabla)\cos(k_z z) g_x(x, k_x(k_z)),$$

$$\underline{\underline{\mathbf{G}}}^{3D}(\mathbf{r}) = -\frac{j\eta}{2\pi k}\int_0^\infty k_\rho dk_\rho (k^2\underline{\underline{\mathbf{I}}} + \nabla\nabla)J_0(k_\rho \rho) g_x(x, k_x(k_\rho)).$$

(23)

Following the procedure leading to Eq. (15), a number $2N_d + 1$ of direct field contributions near the unit cell of interest are extracted and evaluated in closed form. The integration path of the integrals is then deformed to the SDP with pole residues taken into account. Finally, the resulting SDP integrals are regularized as in Sec. 5 and are evaluated using the same quadrature rules as for the scalar PGFs in Eq. (22). Note that, despite the application of the differential operators, the SDP integrands of $\underline{\underline{\mathbf{G}}}^{3D}$ do not have branch cuts and therefore do not require the square-root truncation for the low-frequency regime.

The implementation of the dyadic PGFs allows one, via Eq. (1), to find all the electric-field components resulting from an infinite array of identical dipole sources oriented in any direction. To find the magnetic-field components a similar approach may be used, involving a different dyadic operator but this discussion is omitted for brevity.

## 7. Numerical results

This section presents numerical analysis that shows how the formulations in Secs. 5 and 6 can be used to calculate scalar and dyadic 2D and 3D PGFs. It also studies the convergence rates of the obtained representations and shows that these representations are very efficient for a wide range of structure parameters. The numerical analysis starts with calculating 2D and 3D fields near a 1D array for different structure parameters and proceeds with the study of the error behavior.

Figure 4(a) shows the magnitudes of the scalar 3D PGF $G^{3D}$ and the dyadic components, $G_{xx}^{3D}$ and $G_{yy}^{3D}$, plotted very close to the $x$ axis along the horizontal line $0 < x < 0.5L$ (half of the unit cell is sufficient due to symmetry). The calculations are done using Eq. (15) with $L = 2\lambda$, $y = 0$, $|z| = 10^{-3}L$, and $k_{x0} = (0.85 - 0.001j)k$. The contribution from the source at the zeroth unit cell (i.e. $G_0^{2D,3D}$, $\underline{\underline{\mathbf{G}}}_0^{2D,3D}$) is subtracted to avoid divergence at the unit cell center. Close to the $x$ axis the off-diagonal components of $\underline{\underline{\mathbf{G}}}^{3D}$ vanish, while $G_{yy}^{3D}$ and $G_{zz}^{3D}$ are nearly equal. Physically, this indicates that as $\rho \to 0$ the fields from a periodic array of dipoles oriented parallel to and transverse to the $x$ axis have $TE_x$ and $TM_x$ polarization, respectively. The fields show symmetry in the $y$ and $z$ directions.

Figure 4(b) shows $|G^{2D}|$ plotted along the same line, but for $z = 0$ and $z = 0.1L$. The corresponding dyadic components are not shown since for most 2D problems the fields may be separated into $TE_z$ and $TM_z$ components, each of which may be solved using the scalar PGF $G^{2D}$, so the dyadic operator $\underline{\underline{\mathbf{G}}}^{2D}$ is of limited interest.

In Figs. 5(a) and 5(b) the magnitudes of the 2D and 3D scalar PGF's are plotted along the same line with $L = 0.4\lambda$ and $|z| = 10^{-3}L$ but for different values of $k_{x0}$. Real values of $k_{x0}$ with $k_{x0} < k$ may correspond to the situation of excitation of the array by an incident plane wave, whereas $k_{x0} > k$ may correspond to a situation where an array is excited by a slow waves propagating along the array for which the fields decay away from the array. Complex values of $k_{x0}$, though unphysical, are useful in plane-wave expansions of the fields, and are necessary for finding scattering from such an array.

All results in Figs. 4-5 were obtained using $N_q = 3$, $N_d = 1$, and $\Omega_s^{3D} = 3$, and $\Omega_s^{2D} = 2.5$ ($\Omega_s$ defined in Appendix A determines the integral truncation required for constructing a quadrature rule). These results were compared with results obtained via the conventional spectral representation in Eq. (10) with 80000 summation terms, assumed to be computationally exact. The results were found to be fully convergent (the curves for the representations Eq. (10) and (15) cannot be distinguished visually).

To further demonstrate the rapid convergence of the transverse representation (15) over a wide range of values of the period $L$, Figs. 6(a) and 6(b) show the relative error between the results of the longitudinal and transverse representations obtained via Eqs. (10) and

(15) as a function of the number of quadrature nodes $N_q$ used in the SDP integrations. The error of the transverse representation is defined as

$$\text{Error}^{2D,3D} = \left( \frac{\int_0^{L/2} |G_{\text{exact}}^{2D,3D}(\mathbf{r}) - G_{\text{numeric}}^{2D,3D}(\mathbf{r})|^2 \, dx}{\int_0^{L/2} |G_{\text{exact}}^{2D,3D}(\mathbf{r}) - G_0^{2D,3D}(\mathbf{r})|^2 \, dx} \right)^{1/2}, \quad (24)$$

where $G_{\text{exact}}^{2D,3D}$ are obtained via (10) with a very large $N$ to achieve double-precision accuracy and $G_{\text{numeric}}^{2D,3D}$ are computed using the rapid formulation of (15) with the SDP integrals evaluated via (22) with quadrature rules in Appendix A and B. Note that in the denominator of (24) the zeroth direct term $G_0^{2D,3D}(\mathbf{r})$ is extracted from the reference (numerically "exact") PGF $G_{\text{exact}}^{2D,3D}(\mathbf{r})$ to avoid division by a large number for $\mathbf{r} \to 0$. Without extracting this zeroth term, the resulting error would artificially become significantly lower for small $\mathbf{r}$. Calculations were done using $|z| = \rho = 10^{-3} L$, $k_{x0} = (0.85 - 0.0001j)k$, $N_d = 5$ for three values of periodicity ranging from extremely small ($L = 10^{-6} \lambda$) to moderate ($L = 0.6\lambda$), and to very large ($L = 10\lambda$). Very small periods correspond to dense arrays, such as meshes. Moderate periods (on the order of a wavelength) are characteristic of filters as well as phased array, leaky-wave, and traveling/surface wave antennas. Large periods can also model certain antenna arrays. The parameter $\Omega_s$ was chosen in the range from 2.4 to 6.2. For the 3D PGF, the quadrature rule in Appendix B was used. For the 2D PGF, the quadrature rules in (27) and (32) were used for $L = 10^{-6} \lambda$ and $L = 0.6\lambda$, $L = 10\lambda$, respectively. It is evident that the results converge very rapidly with an increase of the number of integration nodes for any value of the period that would be of practical interest. The components of the dyadic PGFs show similar convergence behavior and, therefore, are not shown.

To demonstrate the effect of poles residing near the integration contour, figure 7(a) shows the 3D PGF calculated with and without pole singularities extracted for $L = 0.4\lambda$, $k_{x0} = 1.05k$, $\rho = 10^{-3} L$, $N_d = 5$, and $\Omega_s = 5.5$. This value of $k_{x0}$ was chosen to be close to $k_0$ such that the presence of the pole affects the rate of the variation of the integrand in (17). Such a value of $k_{x0}$ may correspond to an array excited by a slow wave, e.g. in the case of slow waves traveling on a Yagi-Uda array of dipoles. It is evident that without pole extraction, the results do not converge rapidly. For $k_{x0}$ even closer to $k_0$ the results become absolutely inaccurate. On the other hand, after the pole extraction, a very rapid convergence is obtained.

Next, to show the importance of the square-root truncation of the 2D PGF integrand at low frequencies, figure 7(b) shows the 2D PGF calculated using the non-truncated quadrature rule in (27) and the truncated quadrature rule in (32) with $L = 10^{-3}\lambda$, $k_{x0} = (0.85 - 0.001j)k$, and $|z| = 10^{-3}L$, $N_d = 5$, and $\Omega_s$ ranging from 1.8 to 4.6 for different values of $N_q$. In both cases, the poles were extracted as described in (19)-(22). For this low-frequency regime ($L \ll \lambda$) the presence of the square root for $G_{SDP}^{2D}$ in (19) affects significantly the rate of the integrand variation. As a result, the non-truncated quadrature rule in (27) does not lead to a rapid convergence. For even smaller $L$ the obtained non-truncated results become absolutely inaccurate. However, with the truncated quadrature rule in (32) a rapid convergence is achieved.

Figures 8(a) and 8(b) study the convergence of the transverse representation for a range of observation distances $z$ and $\rho$ from the array for which the Floquet mode representations (10) may still be impractical. The structure parameters are chosen as $L = 0.6\lambda$, $k_{x0} = (0.85 - 0.001j)k$, and $\Omega_s$, fixed for each curve, ranges from 3.8 to 5.5. The error increases away from the array, which is associated with an increased rate of the integrands' variation for larger $z$ and $\rho$. Clearly the largest errors are obtained for the smallest $N_d$ and $N_q$ ($N_d = 3$, $N_q = 6$). The error can be reduced by only a slight increase of $N_d$ and/or $N_q$. It is also important to mention that very low errors with a small $N_d$ and $N_q$ are obtained even for relatively larger displacement from the array ($\rho, z \sim 5\lambda$). This is an important property of the presented formulations as it can allow using the same formulation without switching it with other formulations (e.g. with the Floquet summation in (10)) for a wide range of structure parameters.

## 8. Summary and conclusions

Highly rapid and accurate numerical implementations for scalar and dyadic PGFs for 2D and 3D configurations with 1D (linear) periodicities are introduced. The implementations are based on the expansion of the PGFs in terms of the continuous spectrum integrals over the transverse spectral parameters (Eq. (11)). To achieve a high convergence, the introduced integral representations are cast in a hybrid form in terms of (i) a small number of contributions due to sources located in and around the origin of a periodic unit cell of interest, (ii) a small number of the integrand pole residue contributions corresponding to symmetric combinations of Floquet modes generated by the array, and (iii) the remaining integrals evaluated along the SDP. The integrands of the SDP integrals are regularized by extracting the singular pole behavior near the saddle point of the integrand and integrating it in closed form. The resulting regularized integrands exhibit a slowly varying dependence of the integration variable, which allows implementing highly efficient quadrature rules for the SDP integral evaluation. The integrals for the 2D case are further regularized by truncating the square-root behavior at low frequencies. Numerical simulations were executed to demonstrate strengths of the introduced approach. It was shown that the relative accuracy of 0.1% can be achieved with only about total number of 8 summation components. Double-precision accuracy can be achieved for a total number of about 25 summation components.

It should be noted that the presented schemes for the PGFs reduce to Veysoglu's transformation approach [1, 2, 6, 17] for $N_d = 0$, real $k_{x0}$, and without pole extraction or square-root truncation. However, Veysoglu's approach cannot be used for general complex $k_{x0}$ and cannot be defined on lower Riemann sheets because it does not explicitly take into account the pole residues as done in Eqs. (13) and (15). In addition, due to the singular behavior of the integrand, the Veysoglu approach becomes slowly convergent in the low-frequency regime and for the important case of the "blind" angle regime, where one of the Floquet wavenumbers is near the free-space wavenumber. The schemes presented in this paper efficiently resolve all the potential problems of the Veysoglu and other alternative approaches as explained further next.

1. *Low-and high-frequency regimes:* The presented scheme works seamlessly in the low-frequency regime with $L \ll \lambda$, intermediate frequency regime with $L \sim \lambda$, and high-frequency regime with $L \gg \lambda$. For example, in Fig. 6 the convergence rates for $L = 10^{-6}\lambda$ and $L = 10\lambda$ were nearly identical. This is achieved due to the pole singularity extraction in (19), which makes the integrand vary slowly for any $L$, and square-root truncation in (32), which allows constructing an efficient quadrature rule for 2D PGFs in the low-frequency regime.

2. *Robustness and convenience:* The presented schemes require choosing only a few convergence parameters. These include the parameter $\Omega_s$ determining the integration range limit $s_{\max}$, the number of integration nodes $N_q$, the number of directly extracted terms $N_d$, and the parameter $\Omega_p$ determining the radius of the integral's contributing zone, in which the integrand poles are extracted. All these parameters can be chosen only once to assure a sufficient accuracy, independently of particular problem parameters. For example, $\Omega_s = 6$, $N_q = 10$, $N_d = 5$, $\Omega_p = 2$ would lead to the double-precision accuracy for any practical array parameters and observation locations near the array. Furthermore, the presented schemes lead to a very similar performance for 2D and 3D configurations for scalar and dyadic PGFs. The schemes work seamlessly for any location near the structure, whether along the axis (including the origin) or at a few wavelengths from the axis.

3. *Computational complexity:* The presented schemes are highly computationally efficient. The total number of summation terms is explicitly given by $N_{\text{total}} = (2N_d + 1) + N_q + N_p + \tilde{N}_p$ with $N_{\text{total}} = O(1)$ for any practical structure parameters and observation locations. For example, for the convergence parameter choice in item 2 resulting in double-precision accuracy $N_{\text{total}} = 25$ (assuming $N_p = \tilde{N}_p = 2$ required for most practical values of $L$). If lower accuracy is acceptable, $N_{\text{total}}$ can be reduced further considerably. As discussed in Sec. 5, the presented schemes do not require extensive computational time for evaluating various special-function integrals at every observation location.

4. *Singularity at* $\mathbf{r} \to 0$: The presented schemes explicitly extract the field contribution from the source located at the origin of the $n = 0$ unit cell. As a result the singular behavior of the scalar and dyadic PGFs for $\mathbf{r} \to 0$ is identical to that of the scalar and dyadic free-space PGFs. This fact may be of importance for integral equation methods to evaluate the self-term where the testing and basis functions coincide [1].

The presented schemes can be extended to more complicated environments like layered media. For 2D problems these extensions are straightforward: One only adds in Eq. (11) a summation over the discrete spectrum of guided modes. For 3D problems these extensions are somewhat more involved but possible. Extensions can be developed for 3D problems involving 2D periodicities. The presented ideas also can be used to calculate more complicated Green's functions, e.g. Green's functions for an aperiodic source near an infinitely periodic structure. Such extensions will be reported elsewhere. The scalar and dyadic PGFs can be used in many practical problems including the evaluation of antenna radiation patterns, integral equations for periodic structures, and the study of dispersion properties of periodic arrays.

## Acknowledgments

This research was supported by Nanoscale Interdisciplinary Research Teams (NIRT) program, National Science Foundation (NSF) and by DARPA NACHOS program.

## Appendix A: Quadrature rule for the 2D case

For evaluating the 2D PGF in Eqs. (17) and (19), one needs to construct a quadrature rule for integrals of the following form

$$\int_0^\infty f(s) e^{-\alpha s^2} \frac{ds}{\sqrt{s^2 + 2j}} \approx \sum_{q=1}^{N_q} f(s_q^{2D}) w_q^{2D}, \qquad (25)$$

where $s_q^{2D}$ are the quadrature nodes, $w_q^{2D}$ are the corresponding quadrature weights, $N_q$ is the number of the quadrature nodes, and $f(s)$ is an even function of $s$. Note that since the square-root $(s^2 + 2j)^{-1/2}$ is a complex valued function, the quadrature rule definition in (25) results in complex-valued weights. This definition is employed for convenience of the numerical representation for the SDP integration in (22).

Due to the exponential decay of the integrand, the upper integration limit of the integral (25) can be truncated to a value $s_{max}$, at which the integrand has a sufficiently low magnitude. From (33), $s_{max}$ can be chosen as $s_{max} = \Omega_s/\alpha^{1/2}$, where $\Omega_s$ is a parameter of $O(1)$ chosen as $\Omega_s = \sqrt{\log(1/\varepsilon)}$ with $\varepsilon$ being a prescribed error of the integration truncation. For a given function $f(s)$, the accuracy of the developed quadrature rules depends on the parameter $\Omega_s$. This parameter, through $s_{max}$, determines the integral truncation error $\varepsilon$ and the number of quadrature nodes $N_q$, which in turn determines how many expansion functions in a representation of $f(s)$ are integrated exactly (Eqs. (26) and (31)). Large values of $s_{max}$ require larger $N_q$ to achieve a certain prescribed error, while small values of $s_{max}$ result in a large integral truncation error. For a given $N_q$ there is an optimal $\Omega_s$ (i.e. optimal $s_{max}$) that leads to the minimal achievable error. Likewise, for a given $\Omega_s$ there is a minimal $N_q$ that will lead to a minimal error. In practice one may choose $N_q$ and $\Omega_s$ to be slightly greater than necessary to ensure sufficient convergence for any set of structure parameters.

The quadrature rule in (25) is constructed differently in the high-frequency regime ($\alpha \gg 1$) and low-frequency regime ($\alpha \ll 1$).

A. 1 High-frequency regime ($\alpha \gg 1$, i.e. $L \sim \lambda$ or $L \gg \lambda$)

For $\alpha \gg 1$, the square root in (25) varies more slowly than the rest of the integrand. It is therefore assumed to be a slowly varying function that can be expanded over a small number $P$ of expansion functions in the truncated interval $s \in [0, s_{max}]$. Taking into account the symmetry of the integrand, it can be expanded as

$$\frac{f(s)e^{-\alpha s^2}}{\sqrt{s^2 + 2j}} = \sum_{p=0}^{P-1} a_p \cos\left(\frac{p\pi s}{2s_{max}}\right). \tag{26}$$

Based on this expansion a quadrature rule with equally spaced nodes

$$s_q^{2D} = \frac{2s_{\max}(q+1/2)}{(N_q - 1)}; \quad q = 1,\ldots,N_q,$$

$$w_q^{2D} = \frac{(N_q - 1)}{2s_{\max}(q+1/2)} \frac{1}{\sqrt{\left(s_q^{2D}\right)^2 + 2j}} \tag{27}$$

integrates $P = 2N_q$ expansion (cosine) functions exactly. This simple quadrature rule, which leads to very high accuracy and convergence, is used in Eqs. (17) and (19) to evaluate the SDP integral and lead to (22) for the 2D case for the high-frequency regime.

A.2 Low-frequency regime ($\alpha \ll 1$, i.e. $L \ll \lambda$)

In the low-frequency regime the square-root term $(s^2 + 2j)^{-1/2}$ in (25) (and in (19)) for $G_{SDP}^{2D}$ varies much faster than the exponential term $\exp(-\alpha s^2)$ for $s$ near the origin. To expand this square-root function in the truncated integration range ($s \in [0, s_{\max}]$), a large number of cosine functions is required

$$\frac{1}{\sqrt{s^2 + 2j}} = \sum_{p=0}^{\infty} b_p \cos\left(\frac{p\pi s}{2s_{\max}}\right). \tag{28}$$

Therefore, quadrature rules that assume slowly varying integrands (as in Appendix A.1) would lead to a low-frequency breakdown where no accurate solution can be obtained. However, it is noted that the function $f(s)\exp(-\alpha s^2)$ still varies slowly and can be expanded over a small number $P$ of cosine functions

$$f(s)e^{-\alpha s^2} = \sum_{p=0}^{P-1} a_p \cos\left(\frac{p\pi s}{2s_{\max}}\right). \tag{29}$$

Therefore, the result of the integration in (25) will not change if the expansion of $(s^2 + 2j)^{-1/2}$ is truncated to the same number of $P$ terms

$$\bar{h}(s) = \sum_{p=0}^{P-1} b_p \cos\left(\frac{p\pi s}{2s_{\max}}\right), \tag{30}$$

where $\bar{h}(s)$ is the truncated representation of $(s^2 + 2j)^{-1/2}$. The integrand with the truncated function $\bar{h}(s)$ can be expanded over $2P$ cosine functions

$$f(s)\bar{h}(s)e^{-\alpha s^2} = \sum_{p=0}^{2P-1} c_p \cos\left(\frac{p\pi s}{2s_{max}}\right). \tag{31}$$

For such an integrand the following quadrature rule can be used

$$\begin{aligned} s_q^{2D} &= \frac{2s_{max}(q+1/2)}{(N_q - 1)}; \quad q = 1,...,N_q \\ w_q^{2D} &= \frac{(N_q - 1)}{2s_{max}(q+1/2)} \bar{h}(s_q^{2D}) \end{aligned} \tag{32}$$

to integrate exactly $P$ expansion (cosine) functions. This quadrature rule is used in (22) for the low-frequency case ($\alpha \ll 1, L \ll \lambda$). It should be noted that the quadrature rule in (32) is efficient not only for $\alpha \ll 1$ but also for $\alpha \gg 1$. However, for $\alpha \gg 1$ the simple quadrature in (27) leads to a smaller number of quadrature nodes and, therefore, is used to obtain the results in the paper in this case.

## Appendix B: Quadrature rule for the 3D case

For evaluating the 3D PGF in Eqs. (17) and (19), one needs to design a quadrature rule for integrals of the following form

$$\int_0^\infty f(s)e^{-\alpha s^2} s\, ds \approx \sum_{q=1}^{N_q} f(s_q^{3D}) w_q^{3D}, \tag{33}$$

where $s_q^{3D}$ are the quadrature nodes for the 3D case and $w_q^{3D}$ are the corresponding quadrature weights.

As in Appendix A, the upper integration limit of the integral (33) can be truncated to a value $s_{max}$. The integrand in (33) is assumed to be a slowly varying function that can be expanded over a small number $P$ of expansion functions in the truncated interval $s \in [0, s_{max}]$. Taking into account the weighting function $s$, the semi-infinite integration range, and the behavior of the integrand in (33), this integrand excluding the weighting function $s$ can be expanded over an orthogonal set of Bessel functions

$$f(s)e^{-\alpha s^2} = \sum_{p=1}^{P} a_p J_0\left(\frac{\chi_p s}{s_{max}}\right), \qquad (34)$$

where $\chi_p$ is the $p^{th}$ zero of the Bessel function $J_0$. The basis functions $J_0(\chi_p s/s_{max})$ are orthogonal over the range $s \in [0, s_{max}]$ with respect to the weighting function $s$ [22]. Following the conventional approach for deriving a quadrature rule for functions with a known expansion and based on (33) with (34), the nodes $s_q^{3D}$ and weights $w_q^{3D}$ are found by solving the following system of equations

$$\int_0^{s_{max}} J_0\left(\frac{\chi_p s}{s_{max}}\right) s\, ds = \sum_{q=1}^{N_q} J_0\left(\frac{\chi_p s_q^{3D}}{s_{max}}\right) w_q^{3D}, \qquad (35)$$

where, the integral in the left hand side is given in closed form by $\int_0^{s_{max}} J_0(\chi_p s/s_{max}) s\, ds = s_{max}^2 J_1(\chi_p)/\chi_p$. Equations (35) represent a system of non-linear equations that can be solved for $N_q$ nodes $s_q^{3D}$ and $N_q$ weights $w_q^{3D}$ (total $2N_q$ unknowns) using various modifications of the Newton method. Such solutions result in a Gaussian quadrature rule that integrates exactly $P = 2N_q$ expansion (Bessel) functions $J_0(\chi_p s/s_{max})$ of (34). The quadrature rule (33) with nodes and weights obtained from (35) is used in (22) to evaluate the SDP integrals for the 3D case in (19).

Figures

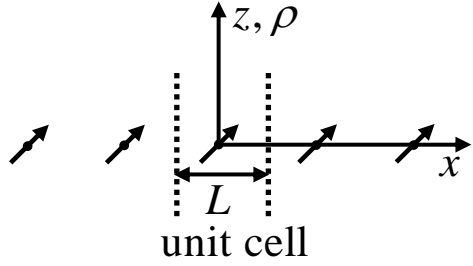

Fig. 1: Structure configuration representing a periodic array of arbitrarily directed dipole sources.

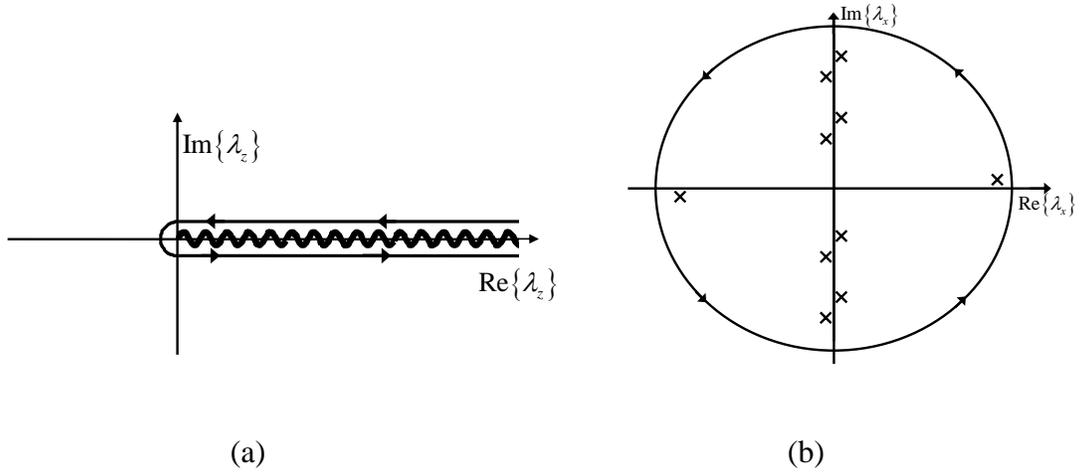

(a)  (b)

Fig. 2: The 1D spectral PGF $g_x$ in the $\lambda_x$ and $\lambda_{z,\rho}$ complex planes. (a) $g_z(z,\lambda_z)$ has a branch cut in the complex $\lambda_z$ plane. The integration around it from Eq. (5) gives the transverse representation in Eq. (11); (b) $g_x(x,\lambda_x)$ has an infinite number of poles in the complex $\lambda_x$ plane. The integration contour in (5) encloses all of them, thus giving the Floquet series in Eq. (10).

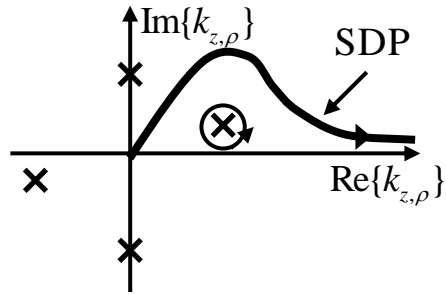

Fig. 3: The complex $k_{z,\rho}$ plane with poles and SDP.

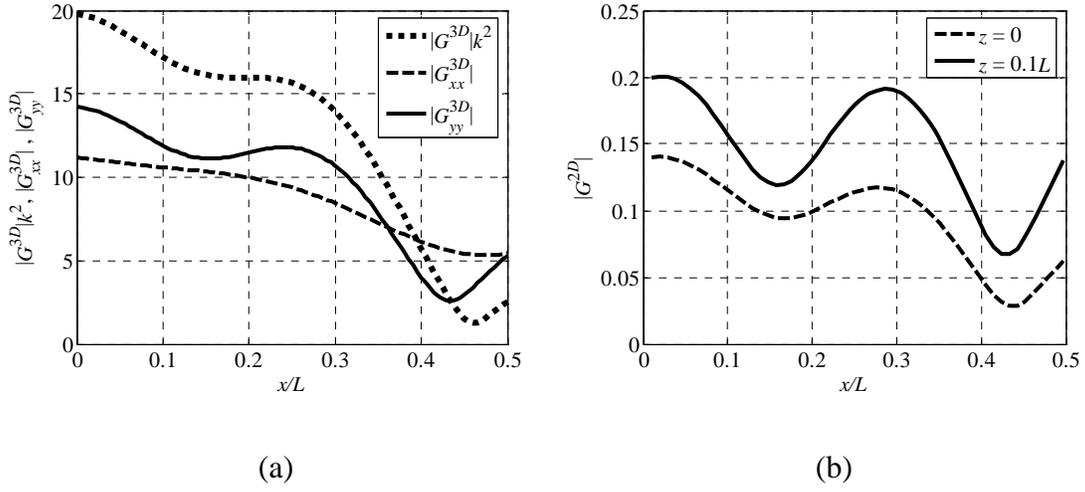

Fig. 4: (a) Magnitudes of the scalar PGF $G^{3D}$ and the dyadic components $G_{xx}^{3D}$ and $G_{yy}^{3D}$ at $|z|=10^{-3}L$, $y=0$ and (b) the magnitude of the scalar PGF $G^{2D}$ for two values of $z$, all plotted along half the unit cell using $L=2\lambda$, $k_{x0}=(0.85-0.001j)k$, $N_q=3$, $N_d=1$, $\Omega_s^{3D}=3$, and $\Omega_s^{2D}=2.5$.

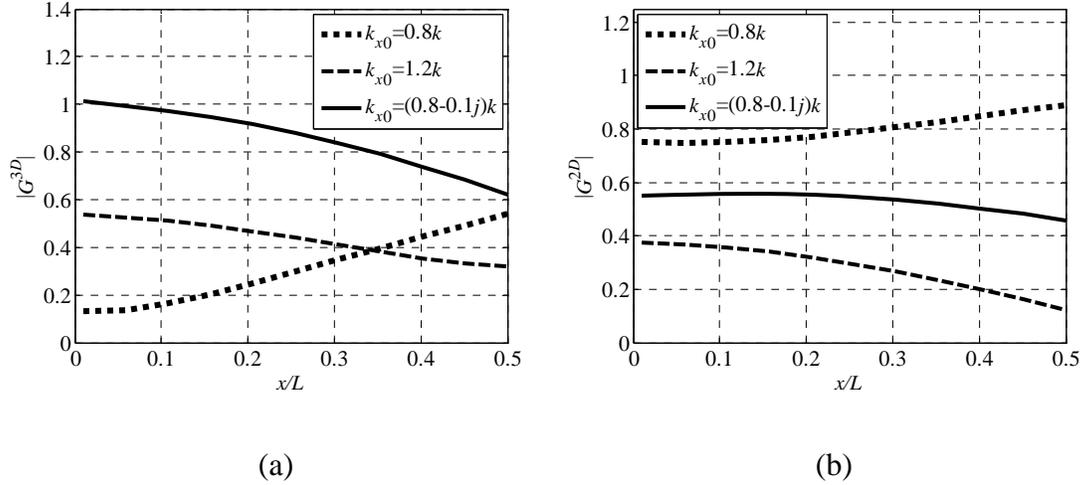

Fig. 5: Magnitudes of the scalar PGFs (a) $G^{3D}$ and (b) $G^{2D}$ plotted along half the unit cell for different values of the linear phase shift parameter $k_{x0}$. Both have $L = 0.4\lambda$, $\rho = |z| = 10^{-3}L$, $N_q = 3$, $N_d = 1$, $\Omega_s^{3D} = 3$, and $\Omega_s^{2D} = 2.5$.

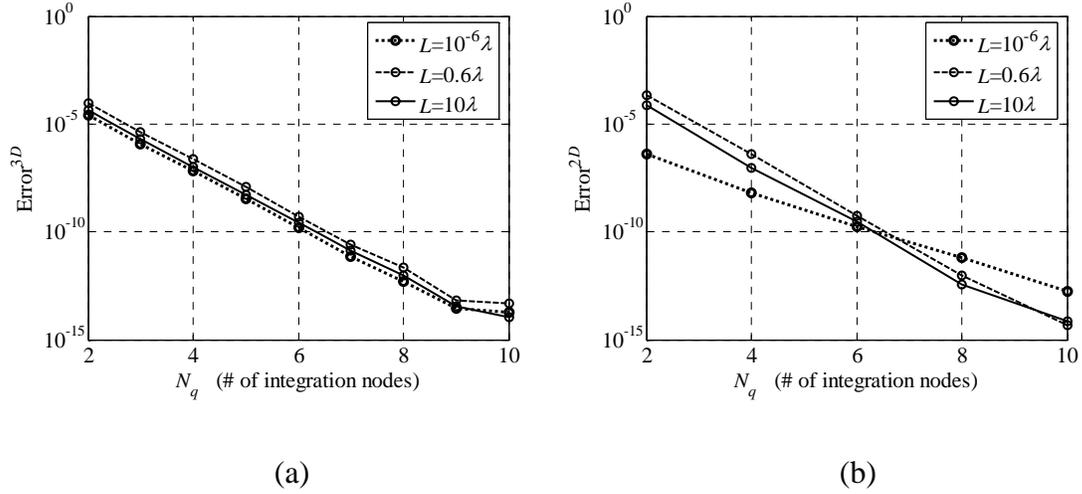

(a)        (b)

Fig. 6: Error of the PGFs (a) $G^{3D}$ and (b) $G^{2D}$ as a function of the number of integration nodes for three values of the spacing $L$, with parameters $\rho = |z| = 10^{-3}L$, $k_{x0} = (0.85 - 0.001j)k$, and $N_d = 5$. The values of $\Omega_s$ vary from 2.4 to 6.2.

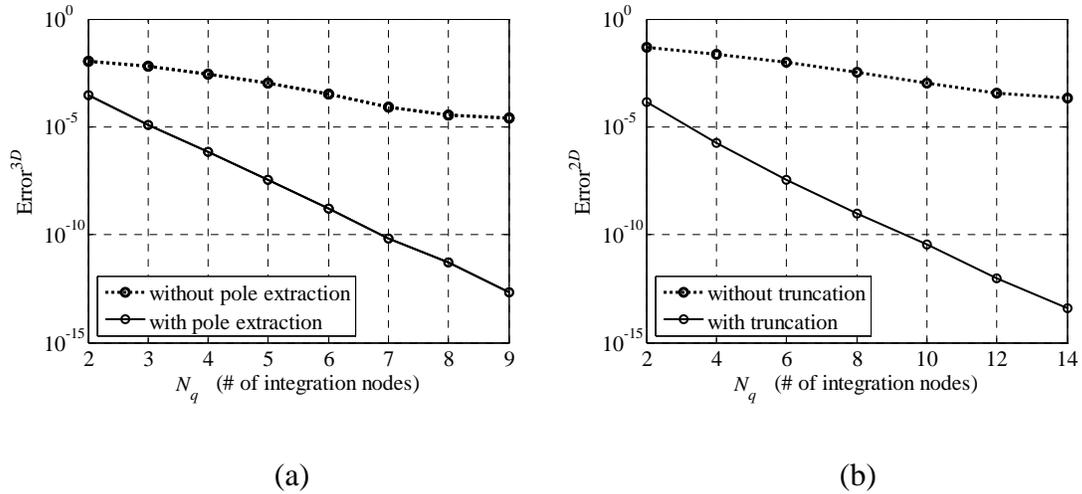

(a)        (b)

Fig. 7: Error of PGFs (a) $G^{3D}$ with and without pole extraction in (19), using $L = 0.4\lambda$, $k_{x0} = 1.05k$, $\rho = 10^{-3}L$, $\Omega_s = 2.6 - 4.6$ with pole extraction, $\Omega_s = 1.8 - 3.6$ without pole

extraction; (b) $G^{2D}$ with and without square-root truncation in (30), using $L = 10^{-3}\lambda$, $k_{x0} = (0.85 - 0.001j)k$, $|z| = 10^{-3}L$, $\Omega_s = 1.6 - 4.6$ with truncation, and $\Omega_s = 0.6 - 2.0$ without truncation.

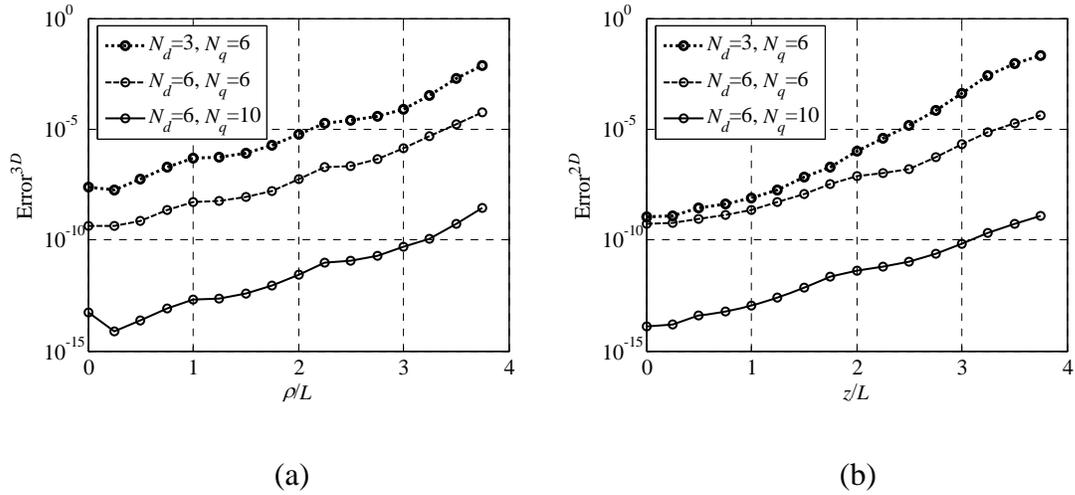

(a)            (b)

Fig. 8: Error of PGFs (a) $G^{3D}$ and (b) $G^{2D}$ as a function of distance from the array for different choices of convergence parameters, using $L = 0.6\lambda$, $k_{x0} = (0.85 - 0.001j)k$, and $\Omega_s$ ranges from 3.8 to 5.5.